\documentclass{ws-procs975x65}

\begin{document}

\title{COSMOLOGICAL BACKREACTION}

\author{Dominik J.~Schwarz\footnote{Talk given at MG12, Paris 2009}}

\address{Fakult\"at f\"ur Physik \\
Universit\"at Bielefeld \\
Postfach 100131 \\
33501 Bielefeld, Germany \\
$^*$E-mail: dschwarz@physik.uni-bielefeld.de}

\begin{abstract}
This work summarises some of the attempts to explain the phenomenon of dark energy as an 
effective description of complex gravitational physics and the proper interpretation of observations. 
Cosmological backreaction has been shown to be relevant for observational (precision) cosmology, nevertheless no convincing explanation of dark energy by means of backreaction has been given 
so far. 
\end{abstract}


\bodymatter

\section{Introduction}

The accelerated expansion of the Universe inferred from the observation of supernovae (SN) Ia
\cite{sn1a} poses one of the biggest puzzles to modern cosmology. While the introduction of a 
simple term --- a cosmological constant --- provides a phenomenological explanation, it leads to 
several conceptual problems at the same time. 

Our minimal model, the inflationary $\Lambda$ cold dark matter model, is very 
successful in describing a large number of cosmological observations like the anisotropies of 
cosmic microwave background (CMB) radiation, the large scale distribution of galaxies, or 
the cosmic expansion. Thus it  is also called concordance model. 
Surprisingly, $95\%$ of the cosmic energy density turn out to be in the form 
of dark components. Despite the success of the model describing the evolution of the Universe, 
it obviously fails to explain its content --- at least unless we uncover the nature of the dark 
components. 

Especially the nature of the phenomenon of ``dark energy'' is completely unclear. There are many 
routes that have to be explored: observational issues, astrophysics issues, testing models of a new component of dark energy, testing modifications to Einstein's equations --- to name just a few popular
directions. Here I focus on the probably most conservative approach to the problem, 
assuming that there are no unknown observational and astrophysical issues.

Accelerated expansion of the Universe seems to start when structures of the size of the 
Hubble scale at matter-radiation equality (the only scale in hierarchical structure formation) 
have grown mildly non-linear. This suggests that dark energy and structure formation might 
be linked in some way. It is also well known that the time evolution of cosmic observables 
and their volume averages do not commute. One might speculate that a mixture of several effects of 
non-linear structure formation and the fact that we actually observe (light-cone) averages 
of local quantities could modify the homogeneous and isotropic ``background'' space-time. 
One often calls that the idea of cosmological backreaction. 

In order to assess the status of these ideas and to motivate their study, I first revisit the key assumptions 
of the concordance model and stress that some pieces of information come entirely from local 
data sets ($z<1$), while some others come from high redshifts at $z > 1000$.  In section 3 the model-independent evidence for cosmic acceleration is discussed, before we briefly turn to some theoretical 
issues related to the concordance model and a sketch of various solutions proposed so far. For the rest of this work I focus on the issue of cosmological backreaction, stressing some of the important issues of cosmic structure formation in section 5 and discussing the importance of averaging in section 5. 
Finally, I summarise some of the open problems.
 
\section{Concordance model} 

The current standard model of cosmology relies four key assumptions: 
\begin{enumerate} 
\item the {\it standard model of particle physics}. It is in excellent agreement 
with experimental fact --- apart from the Higgs, which is searched for at 
Tevatron and LHC experiments. 
\item  {\it general relativity} including a {\it cosmological constant}. Both geodetic motion and
Einstein's field equations are very well tested on various length and time scales. This holds for 
weak fields (Solar system tests) as well as for strong gravitational fields (pulsars), including 
the existence of gravitational radiation. The cosmological constant term can only be probed 
by cosmological observations. It is well motivated by Lovelock's theorem, which tells us that 
adding a cosmological constant to the Einstein-Hilbert action gives rise to the most 
general equation of motion for the space-time metric that is covariant and does not contain 
terms with more than two derivatives. This is a necessary condition for a well posed 
initial value problem. 
\item the idea of {\it cosmological inflation}. It is essential to avoid an enormous amount of 
fine tuning of initial conditions and to make testable predictions on the geometry of the observable 
Universe as well as the statistical properties of distribution of matter and light. If inflation lasts 
for a sufficient number of e-folds, the observable patch of the Universe becomes very close to 
an {\it isotropic and homogeneous space-time with negligible curvature}. Small deviations from 
isotropy and homogeneity due to {\it quantum fluctuations} during inflation are 
distributed in a {\it statistically isotropic and homogeneous} way. As a sufficient
amount of e-folds naturally occurs for inflationary models with a slow-roll epoch, the 
power-spectra of matter and metric fluctuations are close to {\it scale invariant}. In the simplest models 
of cosmological inflation, these fluctuations are also {\it gaussian} distributed and {\it isentropic} 
(entropy per baryon is constant on spatial hypersurfaces).        
\item the existence of {\it dark matter}. There is a plethora of arguments that lead to this conclusion. 
Some of them are astrophysical, like motion in galaxy clusters or galactic rotation curves. Others are
cosmological, like the amount of time necessary for gravitational instabilities to form the observable large scale objects in the Universe. The nature of the dark matter is unknown, candidates are 
weakly interacting massive particles (beyond the standard model of particle physics), primordial 
black holes, invisible axions, \dots An important property of the dark matter is that it must be 
cold, which means that its equation of state must be non-relativistic at the onset of structure formation. 
This property is necessary for sufficiently fast growth of cosmic structures. 
\end{enumerate}
The concordance model has 7 free parameters to be tested by our cleanest probe of cosmology: 
the cosmic microwave background (CMB). These are the photon temperature $T_0$, the 
dimensionless 
energy density of baryons $\omega_{\rm b}$ and of cold matter $\omega_{\rm m}$, the dimensionless Hubble rate $h$, the amplitude of density perturbations $A$, their spectral tilt $n-1$ and the optical 
depth of microwave photons $\tau$. 
The optical depth is not a fundamental parameter, and would follow from the other 
parameters if we had a perfect understanding of astrophysics that leads to the reionisation of the Universe. In the concordance model we have 
\begin{equation}
\Omega_\Lambda = 1 - \frac{\omega_{\rm m}}{h^2}, \quad 
\Omega_{\rm cdm} = \frac{\omega_{\rm m} - \omega_{\rm b}}{h^2}, \quad 
\Omega_{\rm b} = \frac{\omega_{\rm b}}{h^2}, \quad
\Omega_{\nu} = 0. 
\end{equation} 
Since we know form the detection of neutrino oscillations that neutrinos are massive, one could 
actually add $\omega_\nu$ as an 8th free parameter of the concordance model. 

$T_0 = (2.275\pm0.001)$K, obtained by the COBE FIRAS experiment \cite{firas}. 
The remaining 6 parameters can be measured from the CMB  temperature and polarisation 
anisotropies alone, the most recent values have been provided by the WMAP team in their analysis
of 7 years  of data\cite{wmap7yr:Larson}. This is an impressive consistency check of the model.

The addition of massive neutrinos leads to an upper limit on 
$\omega_\nu < 0.014$ (at 95\% C.L.) from the CMB alone\cite{wmap7yr:Komatsu}, 
but at the same time decreases the best-fit value of $\Omega_\Lambda$ by about $10\%$ and 
increases the error bars on $\Omega_\Lambda, \Omega_{\rm cdm}$ and $\Omega_{\rm b}$ by about a 
factor of two. Still, the concordance model seems to be robust against the addition of neutrino
masses.  

We may ask if we can also proof some of the key assumptions. The most prominent one (as it is 
one of the inflationary predictions) probably being the spatial flatness of the Universe. Interestingly 
enough, this does not follow from the CMB alone. We need to add a local measurement. By local 
I mean some probe of the Universe at a redshift $z < 1$. This has been studied in detail by the WMAP team\cite{wmap5yr:Komatsu,wmap7yr:Komatsu}. It has been shown that either a measurement of $H_0$ \cite{H0}, 
baryon acoustic oscillations (BAOs)\cite{BAO} or the Hubble diagram from supernovae (SNe) type Ia
\cite{Union,CfA3,SDSS} can do that. 
Here I would like to stress, that, in contrast to the primary CMB anisotropies, all of those 
observables are based on a sample of objects restricted to a volume much smaller than the local 
Hubble volume. 

Another test is to allow for a general equation of state of dark energy $(w \equiv p/\epsilon \neq -1)$.
As in the case of spatial curvature already for a constant $w$ the CMB alone gives only very weak constraints. Combining the CMB with $H_0$, BAO, SN or a measurement of clusters abundances
confirms $w = -1$ at the $5\%$ and $10\%$ level with and without SNe \cite{wmap7yr:Komatsu}.
Again, we need local ($z < 1$) information to single out the concordance model. 

Thus the question arises if modern galaxy redshift surveys, current SN data sets, as well as recent
cluster catalogues represent fair samples of the Universe and how much they might differ from 
the cosmic mean. Let me stress that no survey of large scale structure has ever covered a volume 
close to the Hubble volume. Typical survey volumes are still well below $V_{\rm H}$, SN surveys are sometimes pencil beams. 

When we add local data sets ($z<1$) to the CMB, more parameters of astrophysical nature are 
needed. To give an example, for nearby supernovae their absolute magnitude $M$ must be 
determined in order to measure the Hubble rate $H_0$. Especially for nearby SNe peculiar 
velocities are important. In order to model them, yet another parameter has to be introduced, 
which is sometimes included into the error bar on the apparent magnitude.   

Adding SNe at larger redshifts ($z \sim 1$) led to the discovery of cosmic 
acceleration\cite{sn1a}. This does not require the independent knowledge of $M$ and $H_0$.
Fitting to models of the cosmic substratum, one only needs a combination of these two parameters,
namely ${\cal M}  = M - 5 \log(H_0) + 25$ --- a quantity that is usually marginalised. Alternatively, one 
can completely avoid ${\cal M}$ by calibrating the SNe with respect to a low redshift sample of SNe.
There is also another hand full of parameters in the light curve fitters, needed to relate the observed SN light curves and spectra to apparent magnitudes in the Hubble diagram. When fitting SNe Ia to the 
concordance model, one typically constrains one parameter, e.g.~$\Omega_\Lambda$. 

A third increasingly important observable are the baryon acoustic oscillations observed in the local 
distribution of galaxies. This feature is the late time reflection of the acoustic oscillations seen in the angular power spectrum of the CMB. It is observed at scales of order $100$ Mpc, thus close to 
the largest scales that present day surveys can probe. BAOs allows us to probe the angular diameter distance at redshifts $\sim 0.1 - 1$, but one should keep in mind that the level of 
sophistication in SN systematics is in advance of BAOs, simply because SNe are being studied since many decades while the BAOs are a very recent subject. As in the case of SNe, some extra 
parameters have to be added to the concordance model, the most prominent one is the bias $b$, but also others encoding 
non-linear corrections to the linear power spectrum are important.  
   
\section{Evidence for accelerated expansion}

The most dramatic consequence of the phenomenon of dark energy, seems to be the 
accelerated expansion of the Universe. Historically, it has been discovered by ruling out the 
Einstein-de Sitter model and finding that the $\Lambda$CDM model provides a much better 
fit to the SN Ia Hubble diagram \cite{sn1a}. These observations fell in line with a problematic age estimate on the basis of the Einstein-de Sitter model and thus the majority of researchers quickly adopted the concordance model. 

However, ruling out the Einstein-de Sitter model is not the same as finding evidence for 
acceleration. Already during the ``supernova revolution'' it was pointed out that an 
inhomogeneous cosmology could describe the Hubble diagram 
\cite{void:age,void:Hubble,Celerier}, match the age limits \cite{void:age}, and 
--- as it turned out later --- even match the angular distance scales of the CMB \cite{void:cmb}.  

It is thus very interesting to try to statistically quantify the evidence for acceleration with as few 
assumptions as possible. The most direct probe of the kinematics of the present Universe are 
SN Hubble diagrams.  In first versions of these tests special parameterizations of the
deceleration parameter $q(z)$ \cite{Turner},
the scale parameter $a(t)$ \cite{Wang} or the Hubble rate $H(z)$ \cite{John} 
have been used. Other published methods are to expand $q$ into principle
components \cite{Shapiro} or to expand the jerk parameter $j$ into a
series of orthonormal functions \cite{Rapetti}.

\begin{figure}
\psfig{file=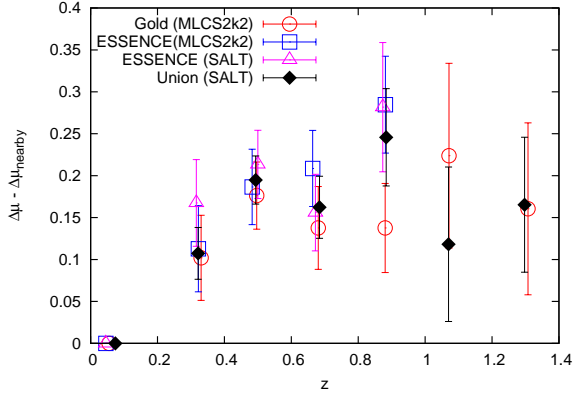,width=3in}
\caption{Cosmic acceleration found at statistically significant levels\cite{Seikel2}. Strong 
evidence at $7.2 \sigma$ significance comes from the Union compilation of SNe using the SALT light curve fitter. This analysis assumes that the Universe is isotropic, homogeneous and spatially flat. Dropping the assumption of spatial flatness allows us provides evidence for acceleration at 
$4.2 \sigma$. The evidence is weaker in case of the MLCS2k2 light curve fitter and for the other 
data sets shown in the figure. Let me stress the importance of the first data bin (at $z < 0.1$), 
which serves as the calibration bin. The error bar of the calibration is taken into account in the 
error bars shown for the bins at higher redshift. \label{figure:acceleration}}
\end{figure}

These assumptions are however not  necessary, if we would like to test the null hypothesis that 
the Universe was always decelerating \cite{Seikel1,Seikel2}. 
Such a model- and  calibration independent
test to quantify the evidence for accelerated expansion still needs to rely on the
assumptions that the Universe is homogeneous and isotropic and that the SNe of the Hubble diagram
are a fair sample of it. The test follows from the simple observation that in a decelerating, flat 
Friedmann model --- independent from its matter content and the validity of Einstein's equations --- 
the luminosity distance satisfies an inequality: 
\begin{equation} 
d_{\rm l}(z)  \leq (1+z) \ln (1+z)/H_0.
\end{equation} 
This translates to the differential distance modulus $\Delta \mu \equiv \mu_{\rm obs} - \mu (q=0) \leq 0$
($q$ is the deceleration parameter), with $\mu \equiv m - M = 5 {\rm log} d_{\rm_l } + 25$.
The resulting test still depends on a combination of $M$ and $H_0$. In order to get rid of this calibration 
dependence we study $\Delta \mu(z) - \Delta \mu(z_{\rm nearby})$. Some results of this analysis are shown in figure \ref{figure:acceleration}. 

Using the Union compilation of SNe Ia\cite{Union} evidence for cosmic acceleration is found at high significance, which is expressed by the fact that all bins are significantly positive, while they would be expected to be negative for a Universe without acceleration. Summing up all bins, the evidence is 
$7.2 \sigma$\cite{Seikel2}. Thus we can conclude that cosmic acceleration is a fact, if we assume isotropy, homogeneity and spatial flatness. The next question is, if we could relax some of these assumptions. Without the assumption of spatial flatness we still found that the evidence for acceleration 
is $4.2 \sigma$. More critical are the assumptions of homogeneity and isotropy. 

\begin{figure}
\psfig{file=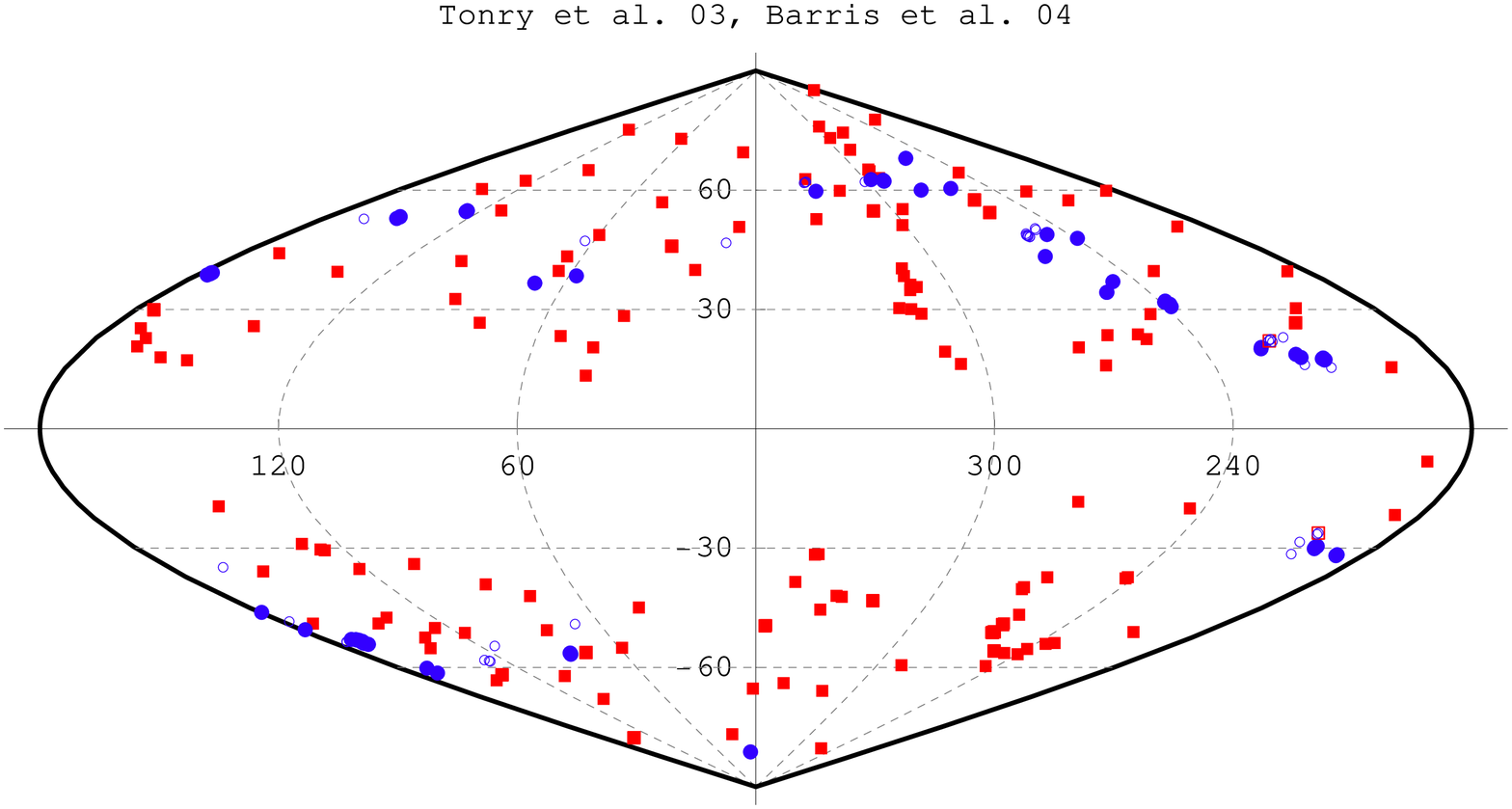,width=2.46in}
\psfig{file=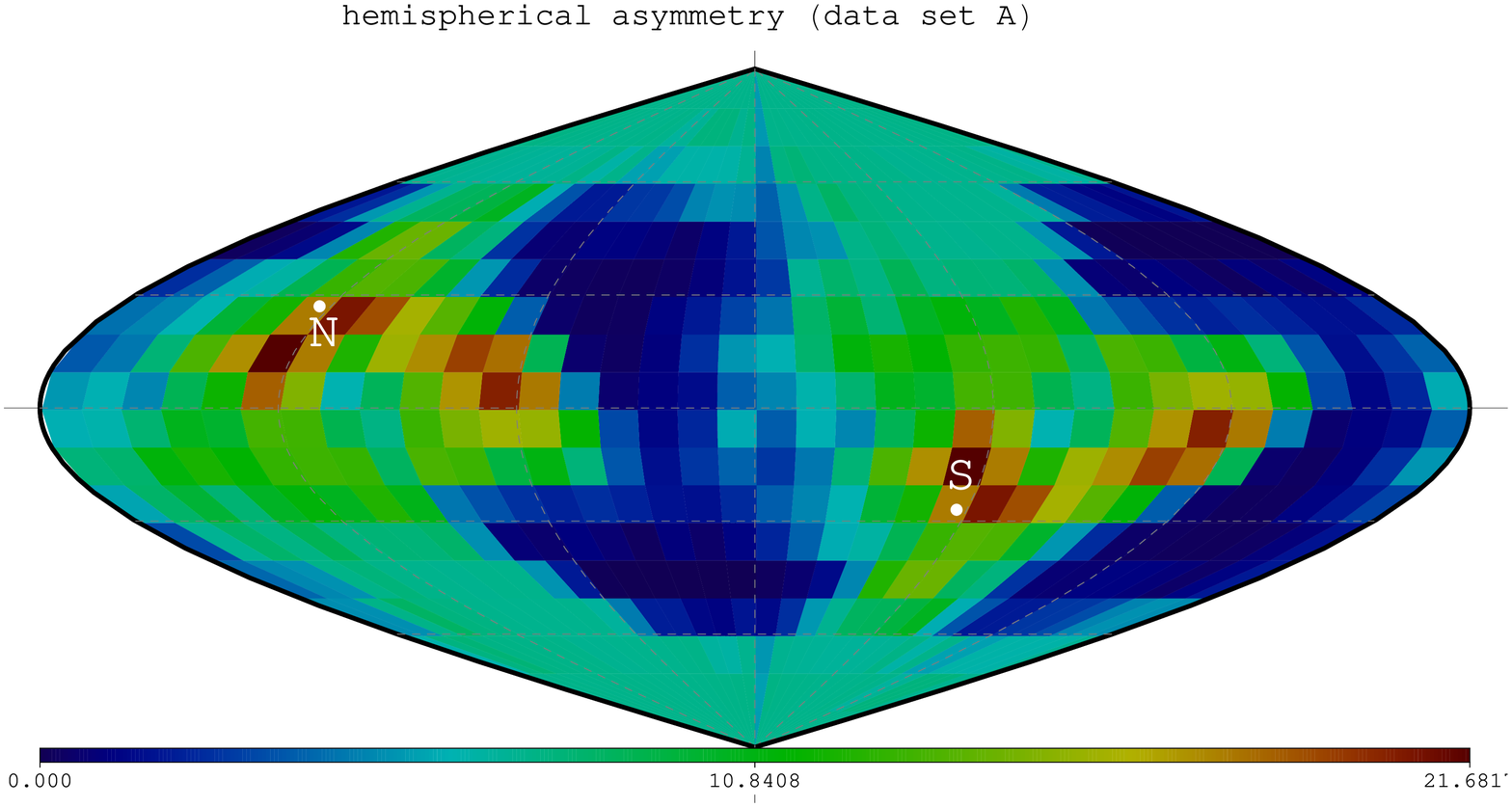,width=2.46in}
\caption{Statistical (an)isotropy of Hubble diagrams from low redshift SNe \cite{weinhorst}. Left panel: 
Distribution of SNe at $z < 0.2$ (red squares) in galactic coordinates. Right panel:  Distribution of 
$\Delta \chi^2$ of Hubble diagrams from pairs of hemispheres, centred on the pixel position shown (galactic coordinates). Red regions show hemispherical asymmetries of the Hubble diagram,  
with $\Delta H_0/H_0 \sim 0.1$, which is statistically significant at the $95\%$C.L. The direction of maximal asymmetry is rather close to the zenith of the equatorial coordinate system, and it is thus 
unclear if this should be regarded as violation of isotropy of local SN data sets or rather as evidence for some systematic issue.}
\label{figure:anisotropy}
\end{figure}

What about isotropy? As we know from the CMB, as well as from radio galaxy surveys and other 
probes of cosmic structure, the deviations from isotropy must be small on large scales. Nevertheless, 
we can test the SN data. It turns out that they are not as isotropic as one would expect 
\cite{mcclure,weinhorst}, but the level of anisotropy found is still consistent with the assumption 
of isotropy \cite{weinhorst}. See figure \ref{figure:anisotropy} for a hemispherical test of SN Ia 
Hubble diagrams and their anisotropy. This analysis found an anisotropy in the value of $H_0$ at 
the $10\%$ level at a statistical significance of $95\%$C.L. Note that a $10\%$ effect in $H_0$ 
could cause the apparent magnitudes to change by $0.2$ mag (c.f. figure \ref{figure:acceleration})!

However, the direction of maximal asymmetry coincides with the zenith of the equatorial 
system, which smells like a systematic issue. Based on other cosmological observations of large scale 
structure, our bias would be that the Hubble flow should be isotropic, maybe apart from some local 
bulk flows.

As already discussed 
above, inhomogeneous cosmologies can fit the SN data, without cosmic acceleration. This does not 
come as a surprise, as e.g.~in the simplest inhomogeneous model --- the spherically symmetric 
Lemaitre-Tolman-Bondi model --- we have two free functions at our disposal. Thus the assumption 
of homogeneity turns out to be absolutely crucial. The homogeneity scale is believed to be at the
$100$ Mpc scale \cite{homogeneity}, or a redshift of $z > 0.1$. We know from observations that 
huge structures exist, like the Sloan Great Wall\cite{Gott}, which seems to extend over $\sim 400$ 
Mpc. 

The most vulnerable point in our argument for cosmic acceleration (see figure \ref{figure:acceleration})
is thus the calibration bin. We can hardly argue that a sample of SNe at $z < 0.1$ can represent a 
fair sample of the large scale distribution. From that point of view, it would be ideal to use 
SNe at slightly higher redshifts to calibrate our test. This will be possible with future data sets\footnote{A preliminary analysis of recent data from SDSS \cite{SDSS}, 
that would in principle allow that test, shows that 
the systematic errors from the two different light curve fitters used by the SDSS team lead to 
conclusions inconsistent with each other\cite{Seikelpriv}.}.  

To summarise this section, under the assumption of isotropy and homogeneity of the Universe at
all scales, we do find model- and calibration independent evidence for cosmic acceleration. 
If we allow for prominent local inhomogeneities, one cannot conclude from SN data alone that 
there is cosmic acceleration.  

\section{Dark physics candidates} 

Although it is remarkable that a model with 7 free parameters is able to fit the majority of 
cosmological observations (it does not fit well at the largest angular scales\cite{large_angles}), 
it has some conceptual problems: 
\begin{description} 
\item {\it Cosmological constant problem}\cite{weinberg}:
If we want to understand if some fundamental constant 
is natural, it is useful to consider dimensionless quantities. There are two fundamental 
constants of gravity, $\kappa = 8 \pi G$ from Newtons constant, and $\Lambda$. Their product is dimensionless and tiny: $\kappa \Lambda \approx 4\times 10^{-120}$. Already without making any reference to vacuum energy density and quantum field theory, this does not seem to be a natural 
choice in the context of the gravitational Lagrangian.
Moreover, quantum field theory should in principle allow us to calculate the energy density of the 
vacuum, but it does not work. Attempts to do so end up with estimates of $\kappa \Lambda \sim 1$, which are obviously wrong. Luckily this is no problem in particle physics, as the vacuum energy 
density would couple to gravity only.
\item {\it Coincidence problem:} This is the question of why $\Omega_{\rm de}(t_0) \sim 
\Omega_{\rm m}(t_0)$, where $t_0$ is the present age or the Universe.  
It seems that the present epoch of the Universe is singled out. However, this argument has an 
anthropic touch as the fact that matter and dark energy are equal 
at some epoch is not at all a surprising result. There is another formulation of the coincidence problem, 
which makes only use of scales intrinsic to the concordance models. 

Hierarchical structure formation is scale free on small scales (apart form a cut-off at the smallest 
scales\cite{ghs}), but knows about the matter-radiation equality scale. That is because growth of structures starts at matter-radiation equality and thus there is a change of slope in the density 
power spectrum and a maximum in the velocity power spectrum at the matter-radiation equality scale. 
In $\Lambda$CDM that scale is at $\sim 100$ Mpc. Non-linear corrections to the linear perturbation 
theory start to become important when the density contrast exceeds $\sim 0.3$\footnote{That is when 
quadratic corrections give rise to $10\%$ effects.}. This typically happens
at $z_{\rm nl}[k = 1/100 \mbox{\ Mpc}] \sim z_{\rm acc} \sim 1$. Thus there is indeed a 
coincidence, which is independent from mankind. 
\end{description}

Many attempts have been made to solve both or one of the mentioned problems. While a solution to the 
cosmological constant problem probably needs an understanding of quantum gravity, which seems to be out of reach at present, the coincidence problem does not necessarily involve physics that is beyond 
our scope. The most prominent alternatives to a {\it cosmological constant} are: 
\begin{description} 
\item {\it Dark energy}\cite{review:de}, 
any component with $p < -\epsilon/3$ at late times. Specific examples are 
quintessence, k-essence, Chaplygin gas and many more. 
\item {\it Modified gravity}\cite{review:mg}. 
Examples are $f(R)$-models, other curvature invariants, non-minimal 
couplings, etc. 
\item {\it Effects of cosmic structure}\cite{Celerier,review:backreaction}. 
This includes ideas that violate the Copernican principle by 
putting us into the centre of the Universe, the idea that the importance of local structures has been 
underestimated, effects of cosmic averaging and non-linear structure formation itself. Cosmological backreaction belongs to this class of ideas, but sticks  to the Copernican principle and to the scenario 
of inflationary cosmology, as well as to the existence of dark matter.   
\end{description} 
In the following I will exclusively focus on what I would call cosmic backreaction, some other 
authors might disagree with my definition of the problem. It seems to me that the most important 
problem is that we often observe averaged quantities, but make our predictions for the local 
quantities. This is the so-called averaging problem. An additional complication is that we actually average over a light-cone volume, never over spatial volumes. Not all observables are volume averages, some are averaged over a surface, some just along the line of sight. 
Now, it is clear that it makes a difference 
whether we first average a quantity, use that average as the initial condition and then evolve 
it in time, or if 
we would start from a complete knowledge of the initial conditions, evolve them in time and average 
at the end of that process to compare to observations. It seems to me that after some intermediate irritation in the community there is no doubt left that averaging and evolution do not commute. 
The dispute is about the order of magnitude of the effect.    

\section{Structure formation}

The already mentioned coincidence of the onset of acceleration and the 
typical scale of large scale structure going non-linear, may provide a hint 
to ask if dark energy is related to structure formation 
\cite{Zimdahl,Schwarz,backreaction,Kolb:2004am}.  

In the concordance model we expect that the first structures are seeded by 
quantum fluctuations during the epoch of inflation. These give rise to acoustic 
oscillations of the primordial plasma with constant amplitude. Cold dark matter, once 
kinetically decoupled from the plasma, starts to grow logarithmically on scales smaller 
than the Hubble horizon during the radiation dominated epoch. Once the Universe becomes 
matter dominated CDM can grow linearly in the scale factor. This scenario is called hierarchical 
structure formation as smallest structures form first and they merge and grow to evolve into larger 
structures. Today scales of $\sim 10$ Mpc have a r.m.s. density 
contrast of order unity (as $\sigma_8 \sim 1$), while the fluctuations at the horizon 
scale ($\sim 4000$ Mpc) are of order $10^{-4}$ in the r.m.s. density contrast. Between these two 
scales there is the matter-equality scale, which due to the onset of linear growth behaviour 
is a distinguished scale in structure formation. In the $\Lambda$CDM model it is at $\sim 100$ 
Mpc. At that scale the density contrast is of order $0.1$, thus non-linear effects start to be important. 
Note that a redshift of $0.1$ corresponds to a distance scale of approximately $400$ Mpc today, 
which corresponds to largest observed structures and can thus not be described by 
linear theory alone. 

Observations and simulations based on Newtonian gravity have revealed that the 
non-linear structure of the Universe forms walls or sheets of matter, which enclose voids,
which might be as big as $100$ Mpc. Within the sheets or at the intersection of sheets 
filaments are formed, which at their densest spots form clusters and superclusters of 
galaxies. 

An interesting scenario is that our local observations are influenced by this cosmic 
space-time structure in a way to mimic dark energy. Instead of studying the idea that we are living 
at the centre of a big over- or underdensity, one can ask what happens if there are many typical 
pronounced over- or underdensities in the Universe, and we happen to live in one of them. 
This idea has attracted a lot of attention recently. The problem is that one is limited to toy models
of space-time, like the swiss-cheese model\cite{swiss_cheese} or wall-void models\cite{wiltshire} 
that are (currently) not derived from first principles.    

The upshot of these studies is that some part of 
the cosmic acceleration seen in SN Hubble diagrams might indeed be due to the effect of a
local inhomogeneity --- although it seems unlikely (see below) that this can explain all of it. 

One might argue that dark energy seems to dominate over the dark matter, but in the light of the idea 
of cosmological backreaction this estimate is wrong. In the spirit of cosmological backreaction
we can view the cosmic substratum as one component with some effective pressure.  The source of 
acceleration within general relativity is the energy density plus three times the pressure. Todays 
deceleration parameter is
\begin{equation} 
q_{\rm eff} = \frac12  \left(1 + 3 \frac{p_{\rm eff}}{\epsilon_{\rm eff}}\right). 
\end{equation} 
If compared with the corresponding relation in the $\Lambda$CDM model,
\begin{equation} 
q_{\rm \Lambda CDM}  = \frac12  \left(1- 3\Omega_\Lambda\right), 
\end{equation} 
we find that $|p_{\rm eff}/\epsilon_{\rm eff}| \sim |\Omega_\Lambda| < 1$. 
Thus, if structure formation should be responsible for the on-set of cosmic acceleration, 
there is hope to use non-linear perturbation theory to understand this mechanism, at least 
at higher redshifts and large scales. 

\section{Cosmic averages} 

Many cosmological observations are averages, e.g.~the
Hubble constant $H_0$. Let us consider an idealised measurement of
$H_0$: One picks $N$ standard candles in a local
physical volume $V$ (at $z \ll 1$), measures their luminosity distances $d_i$ and
recession velocities $v_i=cz_i$, and takes the average $H_0 \equiv
\frac{1}{N}\sum_{i=1}^N \frac{v_i}{d_i}$. In the limit of a very big
sample, it turns into a volume average $H_0 = \frac{1}{V}\int
\frac{v}{d} {\rm d}V$. For objects at $z \ll 1$, the spatial average
is appropriate for the average over the past light cone.

It is still unclear how to best average tensors\cite{hoogen}, a way out is to average only 
scalar quantities. A second problem is to average on the past light cone\cite{coley}.
In the following we restrict our attention to $z \ll 1$, where light-cone averages 
can be replaced by spatial averages. Calculating spatial averages at the Hubble scale and 
beyond is not related to any locally observable quantity. It does however make sense 
to do so at the last scattering surface. 
 
Let me review the formulation of Buchert\cite{Buchert}, which has been used 
to argue that the averaging effect is indeed sizeable and does
give important contributions to cosmology. This set up is well 
adapted to the situation of a real observer.
On large scales, a real observer is comoving with matter, uses her
own clock, and regards space to be time-orthogonal. These conditions
are the definition of the comoving synchronous coordinate system.
Buchert used physically comoving boundaries, which is the most
natural approach in this setup. In the following the Universe is
assumed to be irrotational as a consequence from cosmological
inflation.

In synchronous coordinates, the metric of the inhomogeneous and
anisotropic Universe is $\mbox{d}s^2 = -\mbox{d}t^2 + g_{ij}(t,{\bf
x})\mbox{d}x^i\mbox{d}x^j$, and the spatial average of an observable
$O(t,\bf x)$ in a physically comoving domain $D$ at time $t$ is
defined as
\begin{eqnarray}
\langle O \rangle_D\equiv \frac{1}{V_D(t)}\int_D O(t,{\bf x})
\sqrt{\mbox{det}g_{ij}}\mbox{d}\bf x,\label{average}
\end{eqnarray}
where $V_D(t) \equiv \int_D \sqrt{\mbox{det}g_{ij}} \mbox{d}{\bf x}$
is the volume of the comoving domain $D$. We may introduce an
effective scale factor $a_D$
\begin{eqnarray}
\frac{a_D}{a_{D_0}}\equiv
\left(\frac{V_D}{V_{D_0}}\right)^{1/3}. \label{ad}
\end{eqnarray} 
The effective Hubble
expansion rate is thus defined as $H_D\equiv \dot{a}_D/a_D=\langle
\theta\rangle_D/3$ ($\theta$ being the volume expansion rate).

From the definition (\ref{average}), we obtain effective Friedmann
equations from averaging Einstein's equations,
\begin{eqnarray}
\left(\frac{\dot{a}_D}{a_D}\right)^2= \frac{8\pi G}{3}\epsilon_{\rm
eff}, \qquad -\frac{\ddot{a}_D}{a_D} = \frac{4\pi
G}{3}(\epsilon_{\rm eff}+3p_{\rm eff}), \label{buchert}
\end{eqnarray}
where $\epsilon_{\rm eff}$ and $p_{\rm eff}$ are the energy density
and pressure of an effective fluid,
\begin{eqnarray}
\epsilon_{\rm eff}\equiv\langle\epsilon\rangle_D-\frac{1}{16\pi
G}\left(\langle Q\rangle_D+\langle R \rangle_D\right), \qquad p_{\rm
eff}\equiv-\frac{1}{16\pi G}\left(\langle Q\rangle_D-
\frac{1}{3}\langle R\rangle_D\right). \label{rhoeff}
\end{eqnarray}
$\langle Q\rangle_D\equiv\frac{2}{3}(\langle
\theta^2\rangle_D-\langle \theta\rangle_D^2)-2\langle
\sigma^2\rangle_D$ is the kinematical backreaction ($\sigma^2$ being
the shear scalar), and $\langle R\rangle_D$ the averaged spatial
curvature. They are related by the integrability condition
\begin{eqnarray}
(a_D^6\langle Q\rangle_D)^{^{\textbf{.}}}+a_D^4(a_D^2\langle
R\rangle_D)^{^{\textbf{.}}}=0.\label{integ}
\end{eqnarray}
We define the equation
of state for the effective fluid as $w_{\rm eff}\equiv p_{\rm
eff}/\epsilon_{\rm eff}$. It is highly remarkable that any spatially
averaged dust model can be described by an effective
Friedman-Lema\^itre model.

We can map this effective fluid on a model with dust and ``dark
energy". Let $n$ be the number density of dust particles, and $m$ be
their mass. For any comoving domain $\langle n\rangle_D=\langle
n\rangle_{D_0}(a_{D_0}/a_D)^3$. For a dust Universe, in which
$\epsilon(t, {\bf x}) \equiv mn(t, {\bf x})$, we identify
$\epsilon_{\rm m} \equiv \langle \epsilon\rangle_{D}= m \langle
n\rangle_{D}$, and from (\ref{rhoeff}) the dark energy is hence
described by $\epsilon_{\mathrm{de}}=-(\langle Q\rangle_D+\langle
R\rangle_D)/(16\pi G)$. From (\ref{integ}) we find that constant
$\langle Q\rangle_D=-\langle R\rangle_D/3$ corresponds to the case
of a cosmological constant $\Lambda=\langle Q\rangle_D$. Equations
(\ref{buchert}) and (\ref{rhoeff}) are not closed and additional
input is required. Below we close them by means of cosmological
perturbation theory.

To study the scale dependence of physical observables $\langle
Q\rangle_D$, $\langle R\rangle_D$, $\langle \epsilon\rangle_D$,
$H_D$ and $w_{\mathrm{eff}}$, we calculate them to second order in a
perturbative series of the effective scale factor $a_D$ \cite{Kolb:2004am,LS1,LS2}. 
In synchronous gauge, the linear perturbed metric is
\begin{eqnarray}
\mbox{d}s^2=-\mbox{d}t^2+a^2(t)[(1-2\Psi)\delta_{ij}+D_{ij}\chi]\mbox{d}x^i
\mbox{d}x^j,\nonumber
\end{eqnarray}
where $\Psi$ and $\chi$ are the scalar metric perturbations, $D_{ij}
\equiv \partial_{i}\partial_{j}-\frac{1}{3}\delta_{ij}\Delta$ and
$\Delta$ denotes the Laplace operator in three-dimensional Euclidean
space. The solutions for $\Psi$ and $\chi$ are given in terms of the
time independent peculiar gravitational potential $\varphi({\bf
x})$: $\Psi=\frac{1}{2}\Delta \varphi
t_0^{4/3}t^{2/3}+\frac{5}{3}\varphi$ and $\chi=-3\varphi
t_0^{4/3}t^{2/3}$ (only growing modes are considered)\cite{Kolb:2004am,LS1}.

With the help of the integrability condition, we yield the scale dependence of the averaged physical
observables up to second order \cite{LS2} ($\langle
O\rangle_{D1}\equiv \int_D O {\rm d}{\bf x}/\int_D {\rm d}{\bf x}$
hereafter)
\begin{eqnarray}
\langle Q\rangle_D&=&\frac{a_{D_0}}{a_D}B(\varphi)t^2_0,\label{q2}\\
\langle R\rangle_D&=&\frac{20}{3}\frac{a_{D_0}^2}{a_D^2}\langle
\Delta \varphi\rangle_{D1}-5\frac{a_{D_0}}{a_D}B(\varphi)t^2_0,\label{r2}\\
\langle \epsilon\rangle_D&=&\frac{1}{6\pi G t_0^2}\frac{a_{D_0}^3}{a_D^3},\label{rho2}\\
H_D&=&\frac{2}{3t_0}\frac{a_{D_0}^{3/2}}{a_D^{3/2}}
\left[1-\frac{5}{4}\frac{a_D}{a_{D_0}}t_0^2\langle \Delta
\varphi\rangle_{D1}+\frac{3}{4}\frac{a_D^2}{a_{D_0}^2}t_0^4\left(B(\varphi)
-\frac{25}{24}\langle \Delta \varphi\rangle_{D1}^2\right)\right],\label{theta2}\\
w_{\rm eff}&=&\frac{5}{6}\frac{a_D}{a_{D_0}}t_0^2\langle\Delta
\varphi\rangle_{D1}-\frac{a_D^2}{a_{D_0}^2}t_0^4\left(B(\varphi)-\frac{25}{12}\langle
\Delta \varphi\rangle_{D1}^2\right),\label{w2}
\end{eqnarray}
with $B(\varphi)\equiv \langle
\partial^i(\partial_i\varphi\Delta \varphi)-
\partial^i(\partial_j\varphi\partial^j\partial_i\varphi)\rangle_{D1}
-\frac{2}{3}\langle \Delta \varphi\rangle_{D1}^2$. We see from
(\ref{q2}) -- (\ref{w2}) that these quantities are functions of
surface terms only, so all their information is encoded on the
boundaries. To a first approximation $a_{D_0}/a_D \approx 1 + z$ and we
can see that the perturbative corrections to the kinematic backreaction and to 
the averaged spatial curvature drop much slower than the matter density.
Thus we expect on general grounds of perturbation theory that in a matter dominated 
Universe, averaged curvature and kinematic backreaction will eventually lead to a breakdown 
of perturbation theory. 

\begin{figure}
\psfig{file=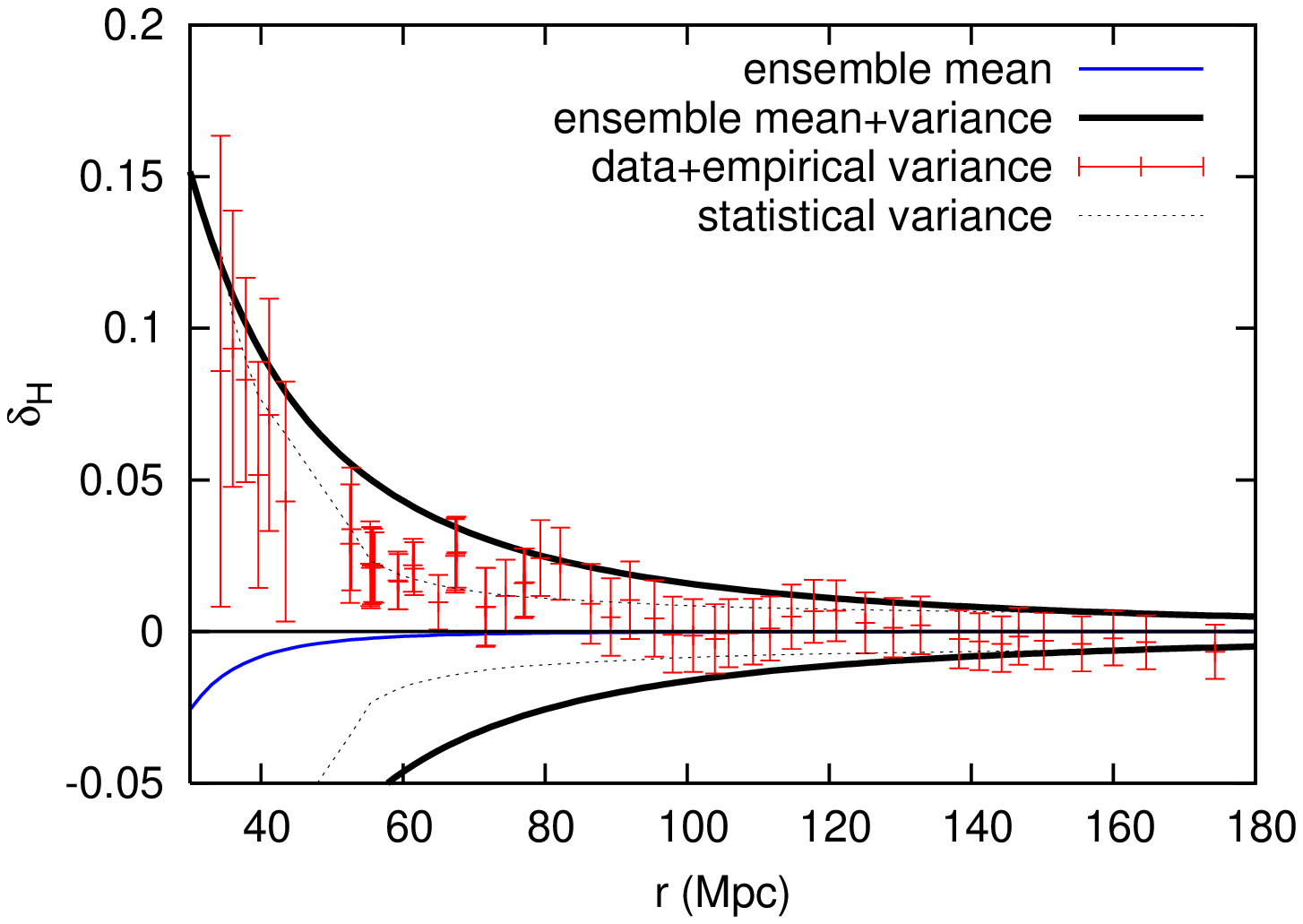,width=2.9in,angle=270}

\psfig{file=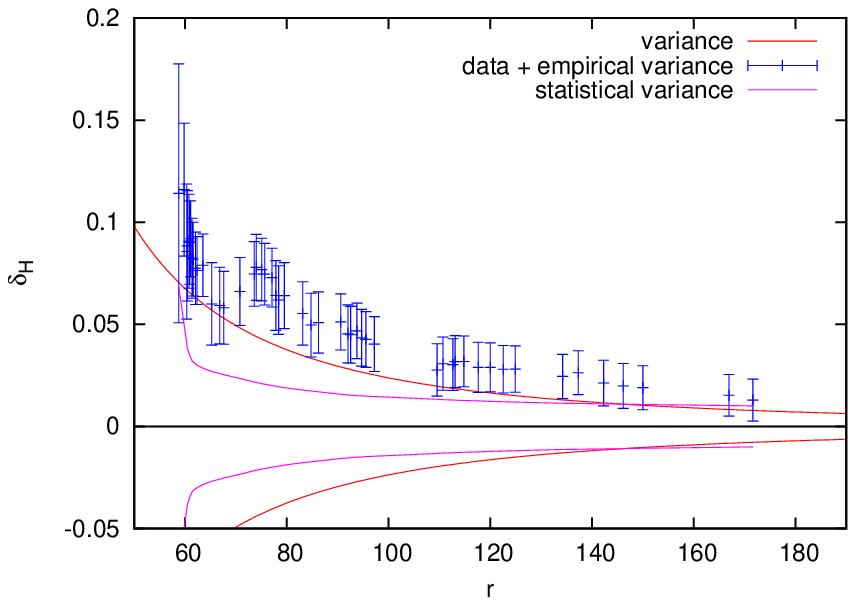,width=4in}
\caption{\label{figure:hubble}Volume and sampling dependence of $\delta_H \equiv (H_D - H_0)/H_0$
of observed Hubble diagrams. The ensemble mean of $\delta_H$ is negligible at scales $> 40$ Mpc (blue curve in the top panel). However the ensemble variance of $\delta_H$ is $5\%$ for a spherical volume with $r \sim 60$ Mpc and falls with $r^{-2}$ (black line in top panel, red lines in bottom panel). The data averages and their empirical variances are shown from two data sets: 
data from the Hubble space telescope key project\cite{Freedman} 
with $H_0 = 72$ km/s/Mpc (top) and the Union 
compilation of SN\cite{Union} with $H_0 = 70.5$ km/s/Mpc and the calibration suggested in Riess et al.\cite{RiessCal} (bottom). In both cases the data are fully consistent 
with the effect of cosmic averaging. While the HST data are also consistent with the hypothesis of pure 
shot noise, $\delta_H(r)$ is well above the expected shot noise between $40$ Mpc and $160$ Mpc for the local SN form the Union set for the assumed calibration and global value of $H_0$. It should be stressed that the latter result suffers from systematic uncertainties regarding the SN calibration. An averaging effect of 5\% in $H_0$ could give rise to a shift of the calibrating bin in figure 1 by $0.1$ mag.}
\end{figure}

One problem is that it is not simple to observe spatial averages as a function of time. A simpler
possibility is to investigate the dependence on the average scale, i.e.\ the sample volume
$V_{D_{0}}^{1/3}$. We show below that cosmological averaging produces
reliable and important modifications to local physical observables, such as $H_0$.

We introduce the quantity\cite{LS2}
\begin{equation}
\delta_H = {H_D - H_0\over H_0},
\end{equation}
where $H_0$ stands for the true global average, which is well defined in the context of perturbation 
theory. $H_D$ is given by (\ref{theta2}). We can now calculate the ensemble average and ensemble 
variance of $\delta_H$. As naively expected the average receives only a contribution form the second 
order perturbations and is thus small, however the variance receives a contribution from the linear 
order and is thus sizeable. We find
\begin{equation} 
\sqrt{{\rm Var}[\delta_H]} \propto \frac{1}{1+z} \left(\frac{r_H}{r}\right)^2 \sqrt{{\cal P_\varphi}}, 
\end{equation} 
which depends on $H_0$ and the dimensionless power spectrum ${\cal P_\varphi}$, which 
is fixed by the CMB temperature anisotropies. This is a unique prediction, which can be 
confronted with data\cite{LS2,SS}, as shown in figure \ref{figure:hubble}.

It seems to me that one can argue that cosmic averaging is expected to give rise to observable effects. 
However, this effect alone cannot explain the accelerated expansion of the Universe. However, 
other effects of averaging have not been quantified yet, such as the effects of averaged curvature.

\section{Open problems and no conclusions}

Probably the most important open problem is to learn how to treat light-cone averages in 
realistic ab initio calculations of the Universe. 

My expectation is that the $100$ Mpc scale is most important for the effects of cosmic 
backreaction and that we have to design tests to decide the issue by 
means of observations. The proposed study of the variance of $\delta_H$ is a first idea in 
that direction. The preliminary conclusion is that cosmic averaging is important, but not 
necessarily an explanation of dark energy.

However, the hardest challenge for cosmological backreaction is to explain why cosmic 
backreaction would mimic $\Lambda$CDM. At the time being, the cosmological constant 
wins the beauty contest. 

\section*{Acknowledgements}

I thank Marina Seikel for interesting and fruitful discussions and collaboration and for providing 
the bottom panel of figure 3. This work was supported by the Deutsche 
Forschungsgemeinschaft (DFG).

\end{document}